\documentclass[conference,a4paper]{IEEEtran}

\usepackage{eso-pic}

\usepackage[french,english]{babel}
\usepackage{amssymb}
\usepackage[T1]{fontenc}
\usepackage{lmodern}
\usepackage{epsfig}
\usepackage{epstopdf}
\usepackage{lastpage}
\usepackage{graphicx}
\usepackage{epstopdf}
\usepackage{lipsum}
\usepackage{amsmath}
\usepackage{mathtools}
\usepackage{times}
\usepackage{ifpdf}
\usepackage{cite}
\usepackage{color} 
\usepackage{slashbox}
\ifCLASSINFOpdf

\fi

\begin{document}

\title{Analysis of the Impact of Impulsive Noise Parameters on BER Performance of OFDM Power-Line Communications}

\author{\IEEEauthorblockN{Kassim Khalil, Patrick Corlay, Fran\c{c}ois-Xavier Coudoux, Marc~G.~Gazalet and Mohamed Gharbi}
\IEEEauthorblockA{IEMN UMR 8520, Department OAE, UVHC, Valenciennes, France\\
Email: kassim.khalil@univ-valenciennes.fr}
}

\AddToShipoutPicture*{\small \sffamily\raisebox{1.8cm}{\hspace{1.8cm}978-1-4244-5997-1/10/\$26.00 \copyright2010 IEEE}}

\maketitle

\begin{abstract}
It is well known that asynchronous impulsive noise is the main source of distortion that drastically affects the power-line communications (PLC) performance. Recently, more realistic models have been proposed in the literature which better fit the physical properties of real impulsive noise. In this paper, we consider a pulse train model and propose a thorough analysis of the impact of impulsive noise parameters, namely impulse width and amplitude as well as inter-arrival time, on the bit error rate (BER) performance of orthogonal frequency division multiplexing (OFDM) broadband PLC. A comparison with the conventional Bernoulli-Gaussian (BG) impulsive noise model exhibits the difference between the two approaches, showing the necessity of more realistic models.

\end{abstract}

\IEEEpeerreviewmaketitle

\section{Introduction}
Power-line communications (PLC) offer nowadays a very interesting alternative to wireless communication systems by reusing the existing electrical network infrastructure to transmit high data rates. However, power line channel suffers from severe conditions such as impulsive noise, channel attenuation, and multipath effects. A good understanding of its characteristics is of great importance when developing PLC transmission chains and simulating the performance of advanced communications technologies.

To overcome multipath effect and channel attenuation, much effort has been dedicated to characterizing and modeling PLC channels \cite{Zimmermann2002,Tonello2007, TonelloTrans2012,Thepaper_MIMO_2014}. Different types of noise are distinguished over PLC networks \cite{ZimmermannNoise2002}, among them the asynchronous impulsive noise is known to be the most detrimental noise term as it exhibits the highest power spectral density (PSD) and may thus cause bit or burst errors in data transmission. It can reach values of 50 dB above the background noise with impulses widths that fluctuate from some microseconds to a few milliseconds \cite{ZimmermannNoise2002}.

Different statistically-based approaches for modeling the PLC asynchronous impulsive noise have been proposed  in the literature \cite{Degardin2002,ModelingNoise2010,TimeAnaly_NoiseAtSource2011}. Bernoulli-Gaussian (BG)~\cite{BGmodel} and Middleton~class~A~\cite{Middleton_ClassA} are two commonly used PLC impulsive noise models, examples of BG noise-based studies are \cite{Ndo_adaptiveNoiseMitigation} and \cite{OFDM_Impulsive_Multipath2005}. These models provide a closed-form and simple probability density function (pdf) expression which is needed in designing optimum receivers of low complexity. However, they do not represent the bursty nature of the impulses observed over PLC channels, as they assume independent impulse emission.  Recently, more realistic models have been proposed which better fit the physical properties of real impulsive noise. These are Markovian-based BG and Middleton noise models, which enable the impulses to be replicated over several consecutive noise samples \cite{MarkovGaussianmodel,MarkovMiddleton2013model}.

In this paper, we derive an easily tractable noise model from the pulse train model proposed in \cite{ZimmermannNoise2002}, which can be fully described by three random parameters: the impulse amplitude, the impulse width and the inter-arrival time, whose statistical properties can be retrieved from measurements \cite{ZimmermannNoise2002}. We propose a thorough analysis of the impact of impulsive noise parameters on the bit error rate (BER) performance of orthogonal frequency division multiplexing (OFDM) broadband PLC.  We applied the conventional BG impulsive noise model and compared the BER results using the two models. The results exhibited the difference between the two approaches, showing the importance of the considered model.

The rest of the paper is organized as follows. The PLC channel model and the impulsive noise model considered in our system are presented in Section~II. Section~III provides the simulation results and  finally Section~IV concludes the paper.


\section{System Model}

\subsection{PLC Channel Model}
An interesting parametric model for PLC channel transfer functions (CTF) based on physical effects, such as multi-path signal propagation and cable losses, is presented in \cite{Zimmermann2002}. The model parameters can be obtained from well-known geometry network measurements. For the purpose of capturing the random aspect of PLC channels and representing its characteristics, statistical extension of this model has been proposed in \cite{Tonello2007}. This work has been refined and built on in \cite{TonelloTrans2012}, where the CTF defined in the frequency band 2-100 MHz is given by: 
\begin{equation}
h(f)=A\sum_{i=1}^{N_{p}}(g_{i}+c_{i}f^{K2})e^{-(a_{0}+a_{1}f^{K})\ell_{i}}e^{-j2\pi f \ell_{i} /\nu },
\label{Eq1}
\end{equation} 
where $A$, $a_{0}$, $a_{1}$, $\nu$, $K$ and  $K2$ are constant parameters and can be obtained  from fitting the model to channel measurements; $a_{0}$, $a_{1}$ and  $K$ describe the characteristics of the cables; $\nu$ is the phase velocity of the wave inside power line cables and equal to $c/\sqrt{\varepsilon_{r}}$, where  $c$  is the speed of light in vacuum and $\varepsilon_{r}$ is the dielectric constant of the insulating material. The parameters $g_{i}$, $c_{i}$, $\ell_{i}$, and $N_{p}$ are random variables  \cite{TonelloTrans2012};  $g_{i}$ and $c_{i}$ are the path gain coefficients that result from the  product of the reflection and transmission coefficients; $\ell_{i}$ is the length of the $i$-th path; $N_{p}$ is the number of paths.

\begin{figure}[!h]
\centering
\includegraphics[scale=0.55]{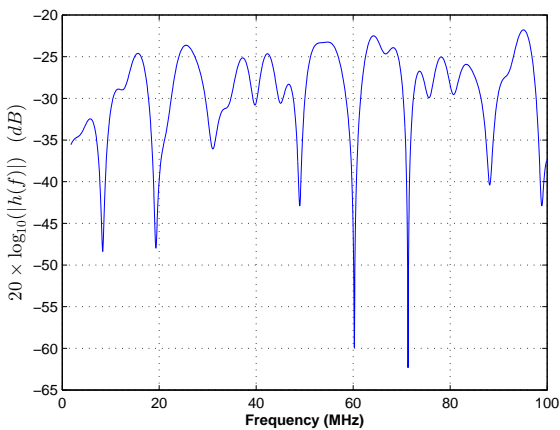}
\caption{Simulated PLC channel transfer function.}
\label{fig1}
\end{figure}

In this work, a channel transfer function $h(f)$ is set from Eq.~(\ref{Eq1}). The chosen values of the path-depend parameters are presented in Table~\ref{table1}; $N_{p}$, $K$, $K2$, $\nu$ and $A$ are $5$, $2.21$, $0.34$, $\frac{2c}{3}$ and  $2.4\times 10^{-5.3}$, respectively. The power of $h(f)$ in dB scale in the frequency band~$1.8-100$ MHz is shown in Figure~\ref{fig1}.

\begin{table}[ht] 
\caption{Parameters for $h(f)$} 
\centering 
\begin{tabular}{c c c c} 
\hline\hline 
$i$ & $g_{i}$ & $c_{i}$ & $\ell_{i}$ \\ [0.5ex] 
\hline 
1 & -0.14 &  0.997 & 5  \\ 
2 &  0.61 & -0.998 & 12 \\ 
3 & -6.61 &  0.998 & 30 \\ 
4 & -0.38 & -0.991 & 35 \\ 
5 & -1.65 & -1.001 & 50 \\ [1ex] 
\hline 
\end{tabular} 
\label{table1} 
\end{table} 

\subsection{Considered Noise Model}

\subsubsection{Bernoulli Gaussian (BG) Model}
One of the popular PLC asynchronous impulsive noise models is the BG model \cite{BGmodel}. According to this model, the total noise seen by the receiver can be considered as a combination of two distributions: an AWGN background noise $w_{k}$ with zero mean and $\sigma_{G}^{2}$ variance, and an independent impulsive noise $i_{k}$ given by
\begin{equation}
i_{k} = b_{k}g_{k},
\end{equation}
where $b_{k}$ is a Bernoulli process and $g_{k}$ is a Gaussian process of variance $\sigma_{I}^{2}$. The total noise, denoted by $n_{k}$, is therefore given by:
\begin{equation}
n_{k}= w_{k} + b_{k}g_{k}
\end{equation} 
and has the following pdf \cite{BGmodel}:
\begin{equation}
p_{BG}(n_{k})=(1-\psi)\,G(n_{k},0,\sigma_{G}^{2})+\psi\,G(n_{k},0,(1+\mu)\sigma_{G}^{2}),
\end{equation}
where $\psi$  is the probability of occurrence of impulsive noise; $\mu=\frac{\sigma_{I}^{2}}{\sigma_{G}^{2}}$ is the impulsive to Gaussian power ratio; $G(\cdot)$ is the Gaussian density defined for real noise as:
\begin{equation}
G(x,m_{x},\sigma_{x}^{2}) = \dfrac{1}{\sigma_{x}\sqrt{2\pi}}\exp(-\dfrac{(x-m_{x})^{2}}{2\sigma_{x}^{2}}).
\end{equation}

\subsubsection{ Bursty Impulsive Noise Model}
In this paper, we derive a low complex noise model from the pulse train model proposed in \cite{ZimmermannNoise2002}. We consider the noise as a sum of background noise and impulsive noise, similar to BG noise model. The general expression of the bursty noise model is given by:
\begin{equation}	
n_{b}(t) = \sigma_{_{G}}\cdot n_{_{G_{1}}}(t)+\sum\limits_{k=1}^{N_{I}} \frac{\sigma_{_{I,k}}}{\sigma_{_{G}}} \mathrm{imp}\left(\dfrac{t-t_{_{A,k}}}{t_{_{w,k}}}\right)\cdot n_{_{G_{2}}}(t),
\label{Eq_burstyModel}
\end{equation}
where $\mathrm{imp(t)}$ is the impulse function with unit amplitude and unit width; $n_{_{G_{1}}}(t)$ and $n_{_{G_{2}}}(t)$  are two zero-mean Gaussian realizations with variance equal to one; $N_{I}$ is the number of impulses present over the transmission time; $\sigma_{_{G}}$ and $\sigma_{_{I}}$ are Gaussian and impulsive noise amplitudes, respectively; $\gamma_{_{k}} =\frac{\sigma_{_{I,k}}}{\sigma_{_{G}}} $, $t_{_{A,k}}$ and $t_{_{w,k}}$ are  impulse amplitude, impulse arrival time and impulse time width, respectively. The three random parameters $\gamma_{_{k}}^{2}$, $t_{_{w,k}}$ and the interarrival time $t_{_{IA,k}}$ \hbox{ ($t_{_{IA,k}}$ = $t_{_{A,k+1}}-t_{_{A,k}} $)} can be statistically modeled in a frequency band up to 20~MHz using an exponential distribution \cite{ZimmermannNoise2002}. Consequently, for a given $\sigma_{_{G}}^{2}$, the bursty noise model can be generated through the three mean parameters of the exponential distributions. The mean parameters of  $\gamma_{_{k}}^{2}$, $t_{_{IA,k}}$ and $t_{_{w,k}}$ are denoted by $\Gamma$, $\lambda$ and $W$, respectively.
 
The impulsive noise is characterized by a high PSD that can reach values of more than 50 dB above the background noise \cite{ZimmermannNoise2002}.  Consequently, three different values of $\Gamma$ were set in this study, which are 10, 100, and 1000. Practical values of $\lambda$ and $W$ in PLC are $ 0.015~s$ and $60~\mu s$, respectively \cite{ZimmermannNoise2002}. Three other values of $\lambda$ of  0.005, 0.05 and 0.1 were also taken into account in order to examine heavier and milder impulsive noise conditions. Another value of $W$ of 1~$\mu s$ was chosen to check the effect of the impulse width characteristic parameter on the system's performance.

\begin{figure}[!h]
\centering
\includegraphics[scale=0.60]{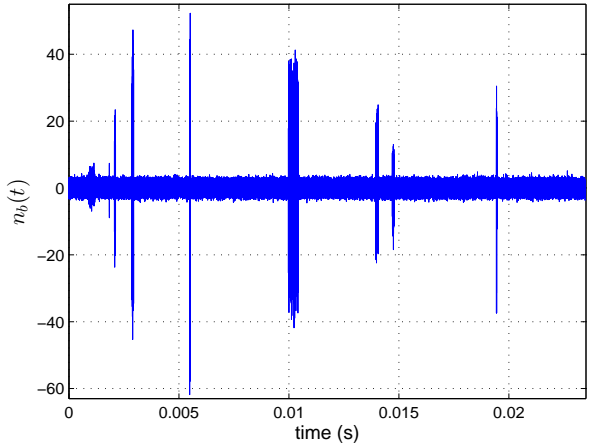}
\caption{Impulsive noise model  with $\lambda=0.005~s$, $\Gamma= 100$ and \hbox{$W = 100~\mu s$}.}
\label{fig2}
\end{figure}

An example of the considered noise model for \hbox{$\lambda=0.005~s$}, \hbox{$\Gamma= 100$} and \hbox{$W = 100~\mu s$} over a time duration of 500 OFDM symbols is presented in Figure~\ref{fig2}. It can be noted that an impulse may affect one or several consecutive OFDM symbols, and many symbols  are not affected by impulsive noise.  It is also important to note that the effect of the impulsive noise is spread to all the subcarriers in an OFDM symbol due to the receiver discrete Fourier transform (DFT) operation. 

\section{Simulation Results}
In this Section, we provide simulation results of BER performance of OFDM system over a PLC channel impaired by impulsive noise. We assumed ideal synchronization and the channel state information is supposed to be known at the receiver. The binary phase-shift keying (BPSK) modulation scheme is used. The OFDM modulator block is implemented using $N=1024$ point inverse fast Fourier transform (IFFT) in the frequency band \hbox{1.8-10 MHz}. The sampling frequency is 24.999936~MHz. The subcarrier spacing is 24.414~kHz, in respect to the subcarrier spacing specified in HomePlug Alliance \cite{HomePlugAV2}. After IFFT, a 150 ($6~\mu s$ in time) length CP is added at the beginning of the OFDM symbol. The length of the extended OFDM symbol is 1174 samples ($46.96~\mu s$ in time). The noise  $ n_{b}(t)$ was generated over a transmission time of $10^{6}$ OFDM symbols before the symbols transmission. During a symbol transmission, the amount of noise from the generated $ n_{b}(t)$ was then added to the symbol. The results are reported as a function of the signal-to-noise ratio (SNR).

\begin{figure}[!h]
\centering
\includegraphics[scale=0.60]{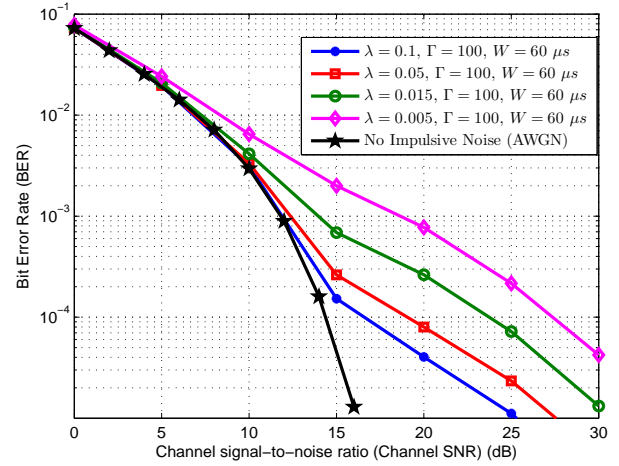}
\caption{BER performance  for different values of $\lambda$ and for $\Gamma=100$ and $W=60~\mu s$.}
\label{fig3}
\end{figure}

The $\lambda$ parameter, which is the mean interarrival time between  impulses, can be considered as the most important characteristic parameter. It gives information about how much the system is affected by impulsive noise. Figure~\ref{fig3} shows its influence on the BER performance for $\Gamma = 100$ and $W=60~\mu s$.  The lower the value of $\lambda$ the more dense the impulses in the system, the more errors produced. This can explain the obtained results where it can be noted that $\lambda$ has a significant influence on the BER performance.

\begin{figure}[!h]
\centering
\includegraphics[scale=0.60]{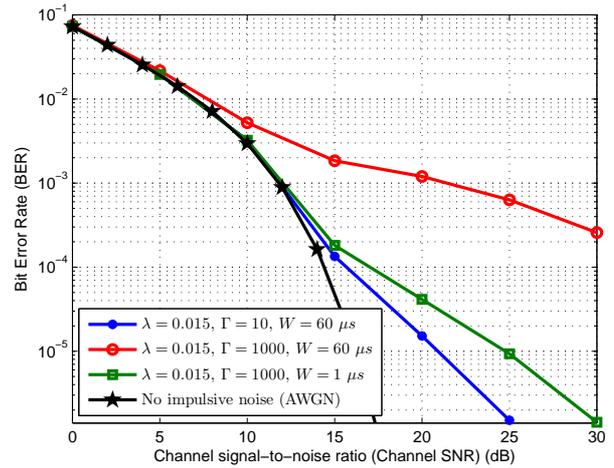}
\caption{BER performance for two different values of each of  $\Gamma$ and $W$ while keeping the other parameters constant.}
\label{fig4}
\end{figure}

In Figure~\ref{fig4}, the system's performance is illustrated for $\lambda=0.015~s$ and for two different values of each of  $\Gamma$ and $W$, while keeping the other parameter constant. This allows us to examine the influence of the two parameters $\Gamma$ and $W$ on the system.  The $\Gamma$ parameter represents the mean ratio of the impulsive noise power to the background noise power. Thus, the higher the $\Gamma$, the more error produced in the system. This can explain the significant difference between the two resultant BERs for $\Gamma=10$ and $\Gamma=1000$ with $W=60~\mu s$.  

The $W$ parameter  represents the mean time width of the impulses, thus the higher the value of $W$ the more signal samples affected by the impulses, the more errors produced. An impulse may affect one or more OFDM symbols resulting in burst errors in data transmission. By comparing the two resultant BERs for $W=60~\mu s$ and $W=1~\mu s$ with $\Gamma=1000$, it can be noted that the $W$ parameter can significantly change the system's performance. At high SNR, a value of $W$ of $1~\mu s$ can affect significantly the system's performance (see Figure~\ref{fig4}).

\begin{figure}[!h]
\centering
\includegraphics[scale=0.60]{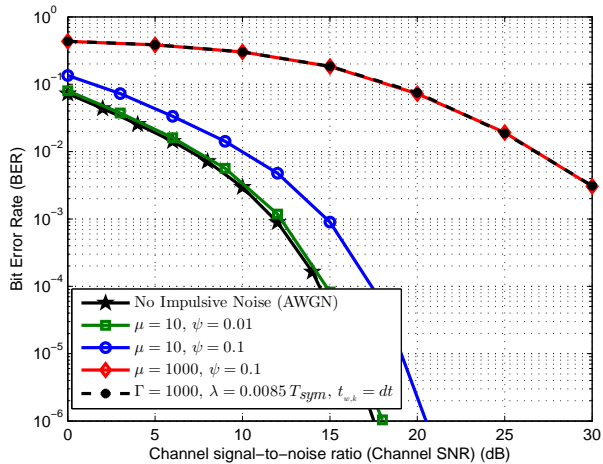}
\caption{BER performance using BG noise model with different values of model's parameters.}
\label{fig5}
\end{figure}

For the reason of comparison, we applied a popular PLC impulsive noise model to the PLC channel, like BG model. The BG model is characterized by two parameters, $\psi$ and $\mu$, as described previously. We checked the effect of the $\psi$ parameter by setting two values of $0.1$ and $0.01$, while keeping $\mu=10$. We also checked the effect of the $\mu$ parameter by setting two values of $10$ and $1000$ with $\psi=0.1$, the results are depicted in Figure~\ref{fig5}. The $\psi$ and $\mu$ parameters are analogous to $\lambda$ and $\Gamma$ parameters, respectively. It can be noted from Figure~\ref{fig5} that they have a significant influence on the system's performance. 

By comparing Figure~\ref{fig5} with  Figures~\ref{fig3}~and~\ref{fig4}, it can be noticed that the resultant BERs using the two noise models follow a different trend. The OFDM performance curves, using the BG model, have the same shape as the BER obtained with AWGN, but with higher power. However, it is not the case, when the considered model is used. This can be attributed to the fact that the characteristic parameters of the two models are different. The $\psi$ parameter represents the probability of the  occurrence of the impulses within the symbol duration. The effect of these impulses is spread over $N$ subcarriers resulting in BER performance that follows the same trend with AWGN performance. On the other hand, by using the considered model, one impulse may affect one or several OFDM symbols, and many symbols may not affected by impulsive noise at all. 

The BG model can be seen as an impulse train model with a similar expression to Eq.~(\ref{Eq_burstyModel}), with the $\mathrm{imp}$ function replaced by the Dirac function (without $t_{_{w,k}}$), and with Bernoulli distributed $t_{_{A,k}}$. In order to reproduce the BG model performance using our considered model, we set the impulses widths $t_{_{w,k}}$ to the sampling time $dt$, that is, we assumed independent impulse emission that are exponentially distributed in time with no memories (widths). The $\lambda$ parameter was chosen to have the same number of impulses within the symbol duration, compared with the BG model, and is given by:
\begin{equation}
\lambda=\dfrac{T_{\textit{symbol}}}{\psi\cdot N},
\end{equation}
where $T_{\textit{symbol}}$ and $N$ are the length of the extended OFDM symbol (with the CP) in time and samples, respectively. For $\mu=1000$ and $\psi=0.1$, $\Gamma=1000$ and $\lambda=0.0085\times T_{\textit{symbol}}$. The result is depicted in Figure~\ref{fig5}, where it can be noticed that they are the same.

\section{Conclusion}
In this paper, we proposed to use a pulse train model that is easily implemented on computer simulations and that well describes the asynchronous impulsive noise over OFDM broadband PLC systems. We examined the influence of the three impulsive noise characteristics, namely the impulse amplitude, the impulse width and the interarrival time on the BER performance. A comparison with the conventional BG impulsive noise model highlighted the difference between the two approaches and showed that the BG model can be considered as a particular case of the considered model.
 
In a future work, the BER performance using the considered model and the one proposed recently in \cite{MarkovMiddleton2013model} will be compared.

\end{document}